\documentclass[aps,twocolumn,bibnotes,amsmath,amssymb,notitlepage]{revtex4-1}

\usepackage{graphicx}
\usepackage{multirow}
\usepackage{color}
\usepackage{units}
\usepackage{mathrsfs}
\usepackage{setspace}
\usepackage[T1]{fontenc}

\usepackage[colorlinks,urlcolor=blue,citecolor=blue,linkcolor=blue]{hyperref}

\usepackage{caption}
\DeclareCaptionLabelSeparator{vline}{ | }
\makeatletter
\def\justified{
  \let\\\@normalcr
  \@rightskip\z@skip \rightskip\@rightskip
  \leftskip\z@skip
  \parindent 0em\relax
  \setlength{\parfillskip}{0pt plus 1fil}}
\DeclareCaptionJustification{justified}{\justified}

\renewcommand{\figurename}{Figure}

\graphicspath{{/}}

\begin{document}

\title{Quantum Chaos in Ultracold Collisions of Erbium}

\author{Albert Frisch}
\author{Michael Mark}
\author{Kiyotaka Aikawa}
\author{Francesca Ferlaino}
\email{francesca.ferlaino@uibk.ac.at}
\thanks{corresponding author}
\affiliation{Institut f\"ur Experimentalphysik, Universit\"at Innsbruck, Technikerstra{\ss}e 25, 6020 Innsbruck, Austria}

\author{John L. Bohn}
\affiliation{JILA, University of Colorado and National Institute of Standards and Technology, Boulder, Colorado 80309-0440, USA}

\author{Constantinos Makrides}
\author{Alexander Petrov}
\altaffiliation{Alternative address: St. Petersburg Nuclear Physics Institute, Gatchina, 188300; Division of Quantum Mechanics, St. Petersburg State University, 198904, Russia.}
\author{Svetlana Kotochigova}
\affiliation{Department of Physics, Temple University, Philadelphia, Pennsylvania 19122, USA}

\date{\today}

\pacs{pacs}

\maketitle

\textbf{Atomic and molecular samples reduced to temperatures below 1 microkelvin, yet still in the gas phase, afford unprecedented energy resolution in probing and manipulating how their constituent particles interact with one another. As a result of this resolution, atoms can be made to scatter resonantly at the experimenter's whim, by precisely controlling the value of a magnetic field \cite{Inouye1998oof}. For simple atoms, such as alkalis, scattering resonances are extremely well-characterized \cite{Chin2010fri}.  However, ultracold physics is now poised to enter a new regime, where far more complex species can be cooled and studied, including magnetic lanthanide atoms and even molecules. For molecules, it has been speculated \cite{Mayle2012sao, Mayle2013sou} that a dense forest of resonances in ultracold collision cross sections will likely express essentially random fluctuations, much as the observed energy spectra of nuclear scattering do \cite{Guhr1998rmt}. According to the Bohigas-Giannoni-Schmit conjecture, these fluctuations would imply chaotic dynamics of the underlying classical motion driving the collision \cite{Bohigas1984coc,Weidenmueller2009rma}. This would provide a paradigm shift in ultracold atomic and molecular physics, necessitating new ways of looking at the fundamental interactions of atoms in this regime, as well as perhaps new chaos-driven states of ultracold matter.}

\textbf{In this report we provide the first experimental demonstration that random spectra are indeed found at ultralow temperatures. In the experiment, an ultracold gas of erbium atoms is shown to exhibit many Fano-Feshbach resonances, for bosons on the order of 3 per gauss. Analysis of their statistics verifies that their distribution of nearest-neighbor spacings is what one would expect from random matrix theory \cite{Brody1973asm}. The density and statistics of these resonances are explained by fully-quantum mechanical scattering calculations that locate their origin in the anisotropy of the atoms' potential energy surface. Our results therefore reveal for the first time chaotic behavior in the native interaction between ultracold atoms.}

In the common perception, atoms are regarded as {\em simple} systems in sharp contrast to {\em complex} molecules, whose behavior is dictated by many (rotational and vibrational) degrees of freedom. The recent realization of dipolar Bose-Einstein condensates and Fermi gases of magnetic lanthanides \cite{Lev2011sdb,Lev2012qdd,Aikawa2012bec,Aikawa2013rfd} made available a novel class of atoms in the ultracold regime. These exotic species, such as erbium (Er), allow to bridge the enormous conceptual gap between {\em simple} atoms and molecules,  potentially providing a natural testbed to explore complex scattering dynamics in a controlled environment. The rich scattering behavior of lanthanides has been pointed out in pioneering experiments at millikelvin temperatures \cite{Hancox2004mto,Connolly2010lsr} and theoretical work on cold collisions of atoms with non-zero angular momenta \cite{Kokoouline2003mcc,Krems2004eia}.

A wealth of intriguing properties in Er, which is the focus of this paper, originates from its exotic electronic configuration. Er is a submerged-shell atom with electron vacancies in the inner anisotropic $4f^{12}$ shell, which lies beneath a filled $6s^2$ shell. As a consequence, it not only has a large magnetic moment of $7$ Bohr magnetons ($\mathrm{\mu_{B}}$) but also has a large electronic orbital (total) angular momentum quantum number of $L=5$ ($J=6$); note that for bosonic (fermionic) isotope the nuclear angular quantum number is $I=0$ ($I=7/2$). Large values for $L$ and $J$ are sources of anisotropy in the interatomic interaction. Moreover, the two-body scattering is controlled by as many as 91 electronic Born-Oppenheimer (BO) interaction potentials, each potential accounting for a specific orientation of $J$ with respect to the internuclear axis. All BO potentials are anisotropic and include at large internuclear separations a strong dipole-dipole interaction (DDI) and anisotropic van der Waals dispersion potentials. This situation is in contrast to that of conventional ultracold atoms, such as alkali-metal atoms, where the scattering is determined mainly by the isotropic singlet and triplet BO potentials \cite{Chin2010fri}. Recent theoretical work predicted the existence of anisotropy-induced Fano-Feshbach resonances in magnetic lanthanides \cite{Petrov2012aif}. This greater complexity brings significant new challenges in understanding and exploiting scattering processes. 

\begin{figure*}[ht]
  \centering
  \includegraphics[scale=1]{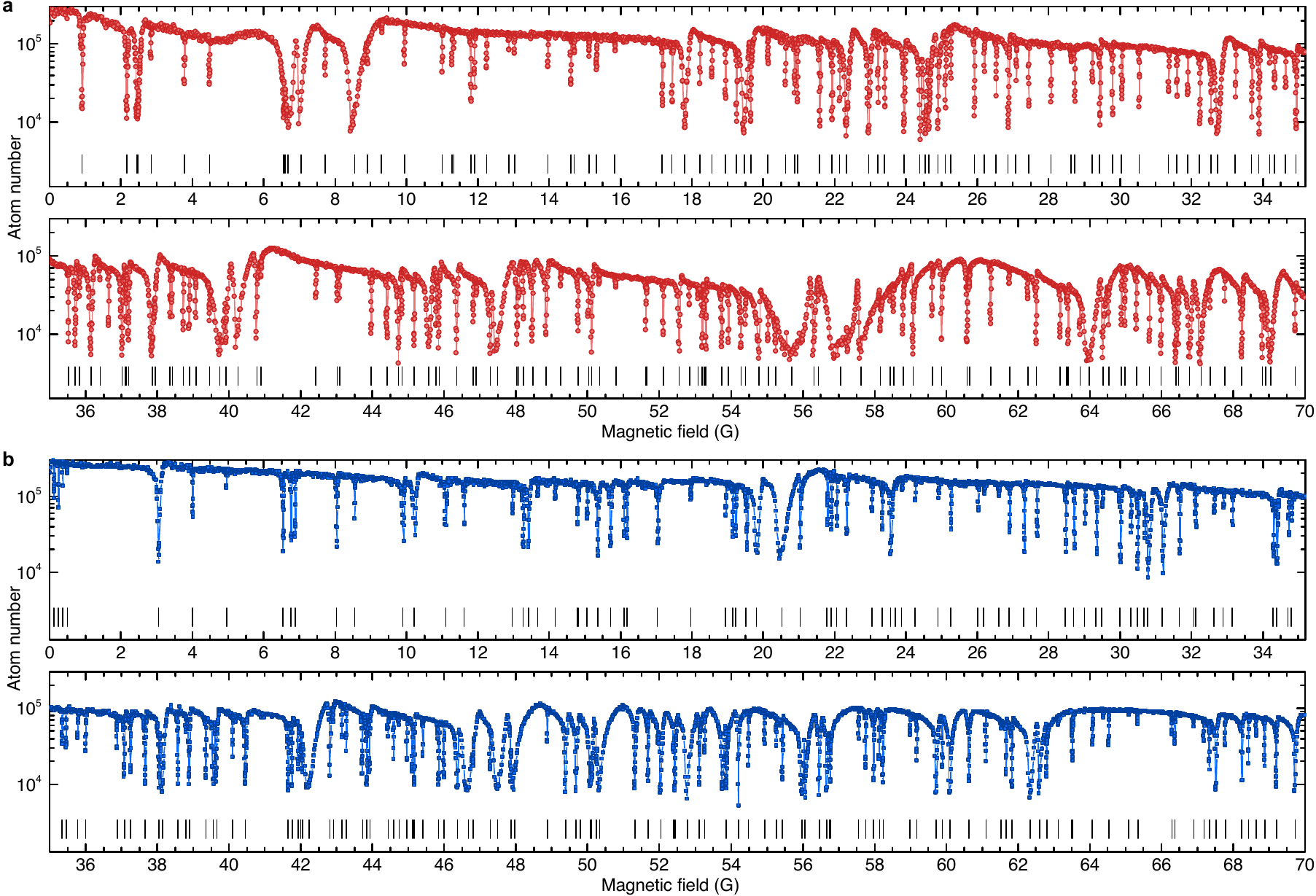}
  \caption{\textbf{Fano-Feshbach spectrum of $^{168}$Er and $^{166}$Er from $0$ to $70\,\mathrm{G}$.} The trap-loss spectroscopy is performed in an optically trapped sample of Er atoms in the energetically lowest Zeeman sublevel $m_{J}=-6$ at a temperature of $330\,\mathrm{nK}$. The atom number is measured after a holding time of $400\,\mathrm{ms}$. \textbf{a}, We observe $190$ Fano-Feshbach resonances for $^{168}$Er and \textbf{b}, $189$ resonances for $^{166}$Er. Resonance positions are extracted by fitting a Gaussian shape to individual loss features; a full list is given in the Supplementary Information.}
  \label{fig:frfeatures}
\end{figure*}

Our experimental study is based on high-resolution trap-loss spectroscopy of Fano-Feshbach resonances in an optically-trapped ultracold sample of Er atoms in the energetically lowest magnetic Zeeman sublevel. We prepare the ultracold sample by following a similar cooling and trapping approach to that described in Ref.\,\cite{Aikawa2012bec} for bosons and Ref.\,\cite{Aikawa2013rfd} for fermions (Method Summary). After the preparation procedure, the ultracold sample typically contains about $10^5$ atoms at a temperature around $400\,\mathrm{nK}$. High-resolution trap-loss spectroscopy consists of many experimental cycles. In each cycle, we ramp the magnetic field to a target value $B$ and hold the atoms for $400\,\mathrm{ms}$ in the optical dipole trap, during which they undergo elastic and inelastic collisions. To probe the loss of atoms from the trap, we record the atom number by applying standard time-of-flight absorption imaging  at zero magnetic field. In the next experimental cycle, we vary the magnetic field value from $0$ to $70\,\mathrm{G}$ with step sizes of a few $\mathrm{mG}$ and repeat the measurement. Figure \ref{fig:frfeatures} shows the loss spectra for $^{168}$Er and $^{166}$Er. For both isotopes, we observe an enormous number of resonant loss features, which we interpret as being caused by Fano-Feshbach resonances \cite{Chin2010fri}. We identify $190$ resonances for $^{168}$Er and $189$ resonances for $^{166}$Er, meaning that we observe about 3 resonances per gauss. We performed similar spectroscopic measurements with the fermionic isotope $^{167}$Er, revealing a much higher density of resonances that exceeds 20 resonances per gauss (Extended Data Fig.\,\ref{figed:spectrfermions}). The fermionic case is complicated by its additional hyperfine structure and detailed studies will be subject of future work.

The immense density of resonances in Er is without precedent in ultracold quantum gases. For comparison, the density  of resonances observed in experiments with ultracold alkali-metal atoms or even mixtures is about two orders of magnitude lower than Er  (c.\,f.\,Ref.\,\cite{Berninger2013frw,Takekoshi2012ttp}). In Er, it is unclear whether a quantitative mapping of the observed resonances is possible at all. In principle there are at least 91 unknown parameters, corresponding to the phase shifts introduced by the BO potentials \cite{Petrov2012aif}. Instead, we focus our theoretical analysis on fundamental questions, such as: Can the observed density of resonances be reproduced by microscopic calculations? Do our results imply the presence of highly anisotropic interactions, which call into play resonant states of high orbital momentum? We answer these questions in the affirmative using full coupled-channel (CC) calculations, supported by an analytical model.

We construct a first-principle CC model for Er+Er scattering to calculate the spectrum of Fano-Feshbach resonances for the experimental conditions. Following Ref.\,\cite{Petrov2012aif}, our model uses the atomic basis set and Hamiltonian (Methods) that includes the radial kinetic and rotational energy operators, the Zeeman interaction, and the 91 anisotropic BO potentials. For small interatomic separations $R$, the BO potentials are calculated using the {\em ab initio} relativistic multi-reference configuration-interaction method \cite{Kotochigova1998crc}. At intermediate to large $R$, the BO potentials are expressed as a sum of multipolar interaction terms. The van der Waals dispersion interaction potentials ($\propto 1/R^6$) are determined from experimental data on atomic transition frequencies and oscillator strengths \cite{nist2013,Lawler2010atp}. An important point is that the dispersion potentials have both isotropic and anisotropic contributions. The latter comes from the non-$S$ state character of the Er electronic ground state. The BO potentials induce thus either isotropic ($\ell$ and $m_\ell$ conserving) or anisotropic ($\ell$ or $m_\ell$ changing) couplings. Here, $\ell$ and $m_\ell$ are the partial wave quantum number and its projection on the magnetic-field quantization axis.

\begin{figure}
\includegraphics[scale=1]{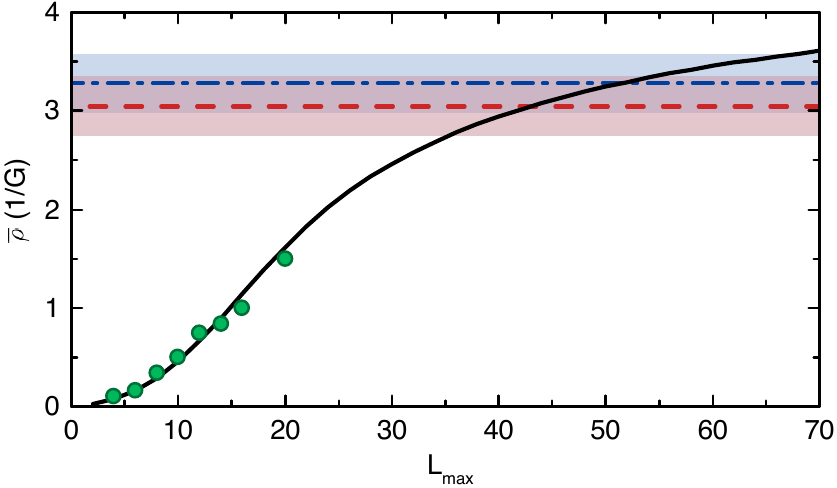}
  \centering
  \caption{\textbf{Mean resonance density for bosonic Er as a function of largest included partial wave $L_{\rm max}$.} CC calculations for  $L_{\rm max}$ up to 20 (circles) and RQDT calculation (solid line) for a magnetic-field region from 0 G to 70 G. For calculations a collision energy of $E/k_{\rm B}=360\,\mathrm{nK}$ is assumed. The mean densities of resonances measured in the experiment are shown for $^{168}$Er (dashed line) and for $^{166}$Er (dash-dotted line) with one-sigma confidence bands (shaded areas).}
  \label{fig:dos}
\end{figure}

We perform CC calculations for bosonic $^{168}$Er, considering $s$-wave ($\ell=0$) collisions and couplings to molecular states with even $\ell$ up to $L_{\rm max}=20$. We calculate the elastic collisional rate coefficient as a function of magnetic field to obtain the Fano-Feshbach resonance spectrum. For $L_{\rm max}=20$, we observe a very dense resonance spectrum with about 1.5 resonances per gauss, which qualitatively reproduces our experimental observation  (Extended Data Fig.\,\ref{figed:CCrate}).
To get deeper insight into the role of the anisotropy of the potentials, we calculate the mean density of resonances $\overline{\rho}$  from our CC calculations for different values of the maximum partial wave $L_{\rm max}$ (Fig.\,\ref{fig:dos}). For $L_{\rm max}$ up to 20, we observe that $\overline{\rho}$ increases with $L_{\rm max}$ in a quadratic manner. This dependence stands in stark contrast to alkali-metal atoms, where high-partial-wave resonances tend to be too narrow to be observed. 

Since our limited computational resources do not allow us to perform calculations for $L_{\rm max}> 20$, it is worth estimating the density of resonances in a simpler way, based on the separated atom quantum numbers \cite{Mayle2012sao}. The key ideas of our model are the following. For each channel $|j_1m_{J,1},j_{2} m_{J,2},\ell m_\ell\rangle$ we define the long-range potential $-C_6/R^6+ \hbar^2\ell(\ell+1)/(2\mu R^2)+ g\mu_{\rm B}(m_{J,1}+m_{J,2})B$, with the isotropic van der Waals $C_6$ coefficient of the BO potentials. Here $\mu$ is the reduced mass, $g$ is the atomic g-factor, and for ground state Er $C_6 = 1723$ a.u.. Fano-Feshbach resonances in our  open $(m_{J,1}\!=\!-6)+(m_{J,2}\!=\!-6)$ channel are due to couplings to the most-weakly bound rovibrational level of closed channels. For a van der Waals potential \cite{Gao2000,Chin2010fri} this bound state has a binding energy that must fall within the $\ell$-dependent energy window $[-\Delta_\ell,0]$ with $\Delta_\ell>0$. The short range potentials are not accurately known and, for each closed channel, there is a probability $dE_b/\Delta_\ell$ of finding a bound state with a binding energy between $E_b$ and $E_b+dE_b$. From Ref.\,\cite{Gao2000} and numerical simulations we find $\Delta_\ell/E_{\rm vdW}\approx38.7+25.5\ell+3.17 \ell^2$, where $E_{\rm vdW}=\hbar^2/(2\mu x^2_{\rm vdW})$ and  $x_{\rm vdW}=\sqrt[4]{2\mu C_6/\hbar^2}/2$. Each closed channel  contributes $g\mu_\mathrm{B}\delta m/\Delta_\ell$ to the mean resonance density, where $g\mu_\mathrm{B}\delta m>0$ is the magnetic-moment difference of the closed and open channels and $\delta m$ is their difference in molecular projection quantum numbers. Adding the contributions for the closed channels  gives $\overline{\rho}$. This counting technique, which we here name random quantum defect theory (RQDT), yields the mean density of states shown in Fig.\,\ref{fig:dos}. For $L_{\rm max}\leqslant 20$, the results of our analytic RQDT agrees very well with the exact CC calculations. For larger $L_{\rm max}$, the density of resonances keeps growing and eventually saturates to a value comparable to the one observed in the experiment. RQDT shows that at least $40$ partial waves need to be considered to reproduce the experimental observations.

Our microscopic models reproduce well the qualitative behavior of the system. However, given the complexity of the scattering, the analysis of ultracold collision data can not  and should not aim anymore at the assignment of individual resonances and the fundamental question of how to tackle complex scattering naturally arises. Historically, spectra of great complexity have been understood within the framework of random matrix theory (RMT), as originally developed by Wigner to describe heavy nuclei containing a very large number of degrees of freedom \cite{Wigner1951oac}. This is an alternative view of the quantum mechanics of complex systems, where individual energy levels and resonances are not theoretically reproduced one-by-one, yet their statistics can be described \cite{Dyson1963sto}. RMT characterizes spectra by fluctuations of their energy levels and classifies their statistical behavior in terms of symmetry classes, e.\,g.\,the Gaussian-orthogonal ensemble (GOE) in the case of a system with time-reversal symmetry, such as neutral atoms.

Following RMT, the distribution of spacings between neighboring levels (or resonances) characterizes the spectral fluctuations of the system and reflects the  absence or the presence of level correlations in terms of a dimensionless parameter, $s$, i.\,e.\,the actual spacing between neighboring levels in units of the mean spacing, $\overline{d}=1/\overline{\rho}$. Whereas the nearest-neighbor spacing (NNS) distribution $P(s)$ of non-interacting levels is Poissonian, $P_{\mathrm{P}}=\mathrm{exp}(-s)$, strongly interacting levels obey a totally different distribution which, in the case of GOE statistics, is known as the Wigner-Dyson (WD) distribution or {\em Wigner surmise} \cite{Dyson1963sto} 
\begin{equation}
\label{PWD}
P_{\mathrm{WD}}=\frac{\pi}{2} s\,\mathrm{exp}(-\pi s^2 /4),
\end{equation}
which shows a strong level repulsion for small $s$, $P_{\mathrm{WD}}(0)=0$. The field of application of the WD distribution is so vast as to make it a universal feature of very complex systems, such as heavy nuclei, disordered conductors, zeros of the Riemann function in number theory, and even risk management models in finance \cite{Guhr1998rmt}. Remarkably, the Bohigas-Giannoni-Schmit conjecture further enriched the field of applications of GOE statistics \cite{Bohigas1984coc}, showing that it applies generally to chaotic systems, such as Rydberg atoms in strong magnetic fields or Sinai billiards, where only few degrees of freedom are relevant, but where motion in these degrees of freedom occurs on a highly anisotropic potential energy surface  \cite{Weidenmueller2009rma}. Recently, it has been speculated that even cold and ultracold atom-molecule collisions will show essential features of GOE statistics \cite{Mayle2012sao,Mayle2013sou}.

\begin{figure} %[ht]
  \centering
  \includegraphics[scale=1]{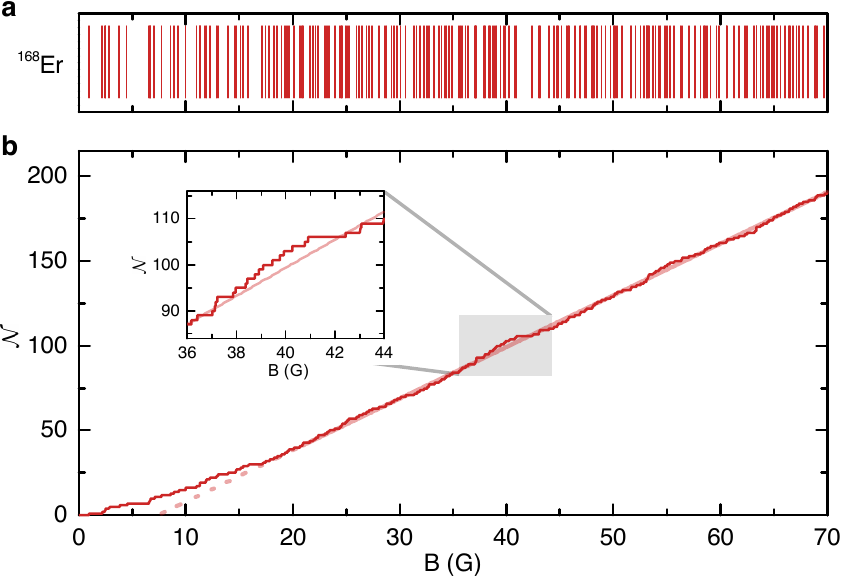}
  \caption{\textbf{Loss-maxima position and staircase function for $^{168}$Er.} \textbf{a}, Positions of the measured loss maxima of Fig.\,\ref{fig:frfeatures} are shown as vertical lines. \textbf{b}, The staircase function shows a linear dependence on the magnetic field at large values. A linear fit to the data above $30\,\mathrm{G}$ is plotted in light colors. The inset shows a magnification of the data to emphasize the step-like nature of the staircase function.}
  \label{fig:staircase}
\end{figure}

Inspired by these works, we statistically analyze both the experimental and calculated Fano-Feshbach spectrum according to RMT. To obtain the NNS distribution of resonances, we first derive $\overline{\rho}$ and the mean spacing between resonances, $\overline{d}$, by constructing the so-called \emph{staircase function} \cite{Weidenmueller2009rma}. This step-like function counts the number of resonances below a magnetic field value $B$ and is defined as $\mathcal{N}(B)=\int\limits_{0}^{B}dB'\, \sum\limits_{i}\delta(B'-B_{i})$, with $\delta$ being the delta function and $B_{i}$ the position of the {\em i}-th resonance. For our experimental data (Fig.\,\ref{fig:staircase}a) the staircase function shows an increase of the number of resonances with $B$, which proceeds linearly at large $B$ and flattens out towards lower magnetic-field values (Fig.\,\ref{fig:staircase}b). The density of resonances is given by the derivative of the staircase function. We evaluate $\overline{\rho}$ in the region above $30\,\mathrm{G}$, where the staircase function shows a linear progression (Supplementary Information) and we obtain $\overline{\rho}=3.0(3)\,\mathrm{G^{-1}}$ and $\overline{d}=0.33(3)\,\mathrm{G}$. We perform a similar analysis with $^{166}$Er and find $\overline{\rho}=3.3(3)\,\mathrm{G^{-1}}$ and $\overline{d}=0.31(3)\,\mathrm{G}$ (Extended Data Fig.\,\ref{figed:statiso}). For CC-calculation data, we find $\overline{\rho}=3.3(3)\,\mathrm{G^{-1}}$ for $L_{\rm max}=20$ (Fig.\,\ref{fig:dos}). We finally derive the NNS distribution for the experimental and CC-calculated data by constructing a histogram of resonance spacings. We choose a number of bins on the order of $\sqrt{N}$, with $N$ being the number of Fano-Feshbach resonances used for analysis \cite{Taylor1997ite}. We then rescale the histogram in terms of the dimensionless quantity $s=B/\overline{d}$ and normalize the distribution in order to obtain $P(s)$. 

\begin{figure} %[ht]
  \centering
  \includegraphics[scale=1]{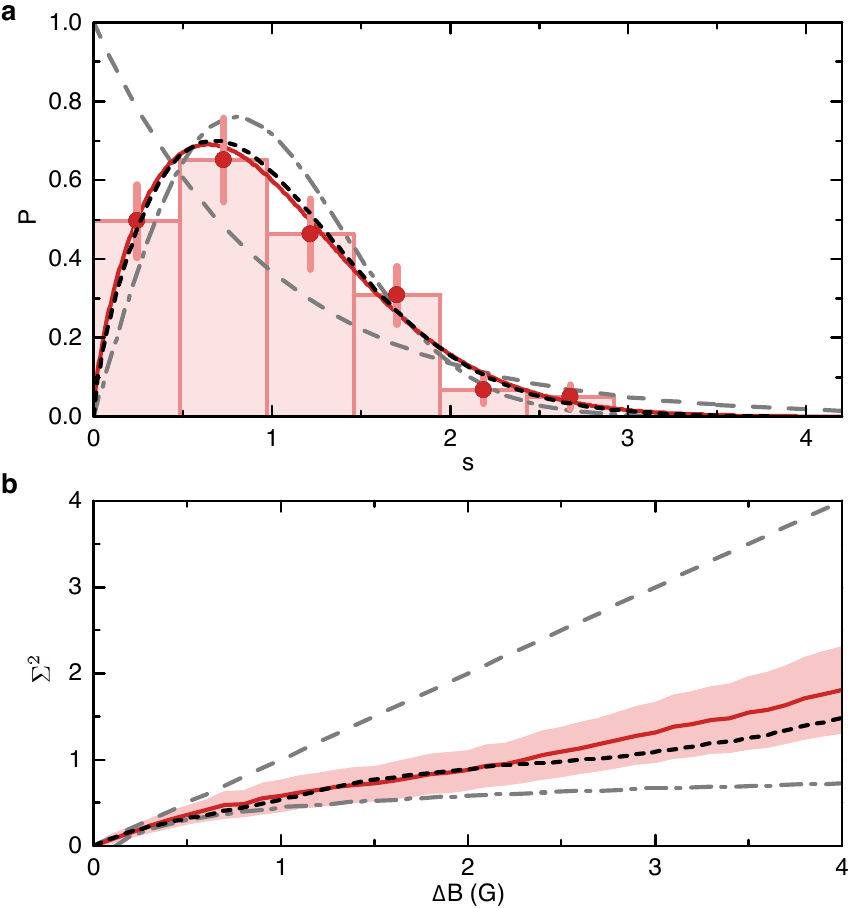}
  \caption{\textbf{NNS distribution and number variance.} \textbf{a}, $^{168}$Er NNS distribution above $30\,\mathrm{G}$ with a bin size of $160\,\mathrm{mG}$.  The plot shows the experimental data (circles) with the corresponding Brody distribution (solid line), the Brody distribution for the CC calculation with $L_{\rm max}=20$ (dotted line), and the parameter free distributions $P_\mathrm{P}$ (dashed line) and $P_\mathrm{WD}$ (short-dashed line). The Brody distribution is given in the Methods section. For the error bars in the experimental data, we assume a Poisson counting error. \textbf{b}, Number variance for the experimental data (solid line) with a two-sigma confidence band (shaded area), the CC-calculation data (dotted line), $P_\mathrm{P}$ (dashed line), and $P_\mathrm{WD}$ (short-dashed line).}
  \label{fig:distribution}
\end{figure}

Figure \ref{fig:distribution} is the main result of our statistical analysis for $^{168}$Er. The plot shows the NNS distribution of the experimental and the CC-calculated Fano-Feshbach resonances together with the parameter-free Poisson and Wigner-Dyson distributions (Eq.\,\ref{PWD}). We see an impressive agreement between the experimental result and the CC calculations. Remarkably, both follow a distribution much closer to the WD than to the Poissonian one. To quantify the agreement with the GOE statistics, we evaluate the reduced chi squared, $\tilde{\chi}^2$, between our data and the Poisson and WD distribution. We find $\tilde{\chi}^2_{\rm WD}=0.9$ and $\tilde{\chi}^2_{\rm P}=2.3$ for our experimental data and $\tilde{\chi}^2_{\rm WD}=0.8$ and $\tilde{\chi}^2_{\rm P}=3.0$ for the data of the CC calculations. The fact that $\tilde{\chi}^2_{\rm WD}\leqslant 1$ confirms that our data are well described by a WD distribution. Similar results are found for $^{166}$Er (Extended Data Fig.\,\ref{figed:distriso}).

To further investigate the spectral correlations, we analyze our data in terms of other statistical quantities, such as the  number variance and the two-gap distribution function (Supplementary Information) \cite{Brody1981rmp}. The number variance $\Sigma^{2}(\Delta B)$ measures the fluctuations of the number of resonances in a magnetic-field interval $\Delta B$ (Methods) \cite{Weidenmueller2009rma} . For non-correlated (Poissonian-distributed) levels,  $\Sigma^{2}=\Delta B$, indicating large fluctuations around a mean value. For quantum chaotic systems, the correlations are strong and the fluctuations are thus less spread out. In this case, $\Sigma^{2} \propto \mathrm{ln}(\Delta B)$. This slower increase of the number variance is regarded as a strong \emph{spectral rigidity} of the system \cite{Weidenmueller2009rma}. Our observations clearly deviate from the Poissonian behavior showing that $\Sigma^{2}$ tends to the WD case (Fig.\,\ref{fig:distribution}b) and confirm the presence of correlations in our system.

To conclude, our observations reproduce the salient features predicted by GOE statistics for chaotic systems, the level repulsion and the spectral rigidity. This implies a degree of complexity in Er+Er cold collisions unprecedented in any previous ultracold scattering system.  Our results bring the powerful analytical tools of quantum chaos to bear \cite{Weidenmueller2009rma}. In particular, these approaches connect the large-scale structure of the spectra to simple features such as the shortest closed classical orbits in the potential energy surface, where these connections are made by the Gutzwiller trace formula \cite{Gutzwiller1990cic}. Identifying the most important closed orbits will then shed light on the potential energy surface itself, providing a route to describing ultracold collisions that is complementary to the elaborate close-coupling calculations that will be difficult to connect in detail with the data.

Erbium represents the first occasion where statistical analyses and chaotic behavior are important to ultracold collisions, but they will not be the last. Specifically, much experimental effort is being exerted toward producing ultracold molecular samples, which also enjoy highly anisotropic potential energy surfaces. Learning to read complex spectra, by acknowledging their essentially chaotic nature, represents a turning point in how the field will consider ultracold collisions in the future and provide new inroads into ultracold chemistry.

\textbf{METHODS SUMMARY}

\textbf{Sample preparation.}
For bosonic sample preparation we follow the approach of Ref.\,\cite{Aikawa2012bec}. We obtain about $3 \times 10^{5}$ optically-trapped atoms at a density of $3 \times 10^{13}\,\mathrm{cm^{-3}}$. The trap-loss spectroscopy is performed in a trap with frequencies of $(\nu_{x},\nu_{y},\nu_{z}) = (65,26,270)\,\mathrm{Hz}$. The temperature of the cloud is measured by time-of-flight imaging  at $0.4\,\mathrm{G}$ and gives $T_{168}=326(4)\,\mathrm{nK}$ and $T_{166}=415(4)\,\mathrm{nK}$, respectively. We ramp the magnetic field within $10\,\mathrm{ms}$ to a probe value between $0$ and $70\,\mathrm{G}$, and hold the atomic cloud for $400\,\mathrm{ms}$ in the optical dipole trap. We observe an increase of the temperature up to $560\,\mathrm{nK}$ at a magnetic field of about $50\,\mathrm{G}$ due to the ramping over many Fano-Feshbach resonances.
For fermionic sample preparation we follow the approach of Ref.\,\cite{Aikawa2013rfd}. We obtain about $1.2 \times 10^{5}$ fermionic atoms at a density of $2 \times 10^{14}\,\mathrm{cm^{-3}}$ and at a temperature of $0.4\,T_{\rm F}$, where $T_\mathrm{F}=1.0(1)\,\mathrm{\mu K}$ is the Fermi temperature. The trap frequencies are $(427,66,457)\,\mathrm{Hz}$.

%\textbf{Full Methods} and any associated references are available in the online version of the paper at www.nature.com/nature.

%\bibliography{ultracold,statFR}

%

%\textbf{Supplementary Information} is linked to the online version of the paper at www.nature.com/nature.

\hspace{10pt}

\textbf{Author contributions} \newline
A.F., M.M, K.A., and F.F. did the experimental work and statistical analysis of the data, C.M., A.P., and S.K. did the theoretical work on CC calculations and RQDT, J.L.B. did the theoretical work on RMT. The manuscript was written with substantial contributions from all authors.

\label{acknowledgments} 
\textbf{Acknowledgements} \newline
The Innsbruck group thanks R.\,Grimm for inspiring discussions and S.\,Baier, C.\,Ravensbergen, and M.\,Brownutt for careful reading of the manuscript. S.\,K.\,and A.\,P.\,thank E.\,Tiesinga for useful discussions. J.\,L.\,B.\,is supported by an ARO MURI. The Innsbruck team is supported by the  Austrian Science Fund (FWF) through a START grant under Project No.\,Y479-N20 and by the European Research Council under Project No.\,259435. K.\,A.\,is supported within the Lise-Meitner program of the FWF. Research at Temple University is supported by AFOSR and NSF PHY-1308573.

\newpage

\section*{METHODS}

\textbf{Experimental procedures.}
For bosonic sample preparation we follow the approach of Ref.\,\cite{Aikawa2012bec}. In brief, after the magneto-optical trap \cite{Frisch2012nlm}, we load the atoms in an optical dipole trap composed of two laser beams in horizontal ($1{,}064\,\mathrm{nm}$, $0.4\,\mathrm{W}$, single-mode) and vertical direction ($1{,}064\,\mathrm{nm}$, $4.0\,\mathrm{W}$, broadband Yb fiber-laser). In the trap, we force evaporation by ramping down the power of both trapping laser beams within $6.2\,\mathrm{s}$, in the presence of a homogeneous magnetic field of $0.4\,\mathrm{G}$ to prevent spin-flips to other Zeeman states. We stop evaporative cooling before the onset of Bose-Einstein condensation. Our final trap has frequencies of $(\nu_{x},\nu_{y},\nu_{z}) = (65,26,270)\,\mathrm{Hz}$ and contains about $3 \times 10^{5}$ atoms at a density of $3 \times 10^{13}\,\mathrm{cm^{-3}}$. The temperature of the atomic cloud is measured by time-of-flight imaging for both isotopes at $0.4\,\mathrm{G}$ and gives $T_{168}=326(4)\,\mathrm{nK}$ and $T_{166}=415(4)\,\mathrm{nK}$. We ramp the homogeneous magnetic probe field up to $70\,\mathrm{G}$ within $10\,\mathrm{ms}$ and hold the atomic cloud for $400\,\mathrm{ms}$ in the optical dipole trap. The magnetic field is suddenly ($<5\,\mathrm{ms}$, limited by eddy currents) switched off and the atom number and size of the cloud is measured via absorption imaging after a time of flight of $15\,\mathrm{ms}$. We observe an increase of the temperature up to $560\,\mathrm{nK}$ at a magnetic field of about $50\,\mathrm{G}$ due to ramping over many Fano-Feshbach resonances. For fermionic sample preparation we follow the approach of Ref.\,\cite{Aikawa2013rfd}. We obtain about $1.2 \times 10^{5}$ fermionic atoms at a density of $2 \times 10^{14}\,\mathrm{cm^{-3}}$ and at a temperature of $0.4\,T_{\rm F}$, where $T_\mathrm{F}=1.0(1)\,\mathrm{\mu K}$ is the Fermi temperature. The trap frequencies are $(427,66,457)\,\mathrm{Hz}$.

\textbf{Magnetic-field control.}
An analog feedback loop stabilizes the current for the homogeneous magnetic-field coils with a relative short-term stability of better than $2 \times 10^{-4}$. Calibration of the magnetic field is done by driving a radio-frequency transition between Zeeman states $m_{J}=-6$ and $m_{J}=-5$. Trap-loss spectroscopy is carried out in steps of $20\,\mathrm{mG}$ (out of resonance) and $5\,\mathrm{mG}$ (on resonance). The long-term offset stability of the magnetic field was observed during the data recording period to be better than $4\,\mathrm{mG}$ within one week.

\textbf{Coupled-channel calculations.}
We perform exact CC calculations for Er+Er scattering in the basis $|j_1m_{J,1},j_2 m_{J,2},\ell m_\ell\rangle$ \allowbreak $\equiv$ \allowbreak $Y_{\ell m_\ell}(\theta,\phi)|j_1m_{J,1}\rangle |j_2 m_{J,2}\rangle$, where $\vec{j}_{a=1,2}$ are the atomic angular momenta with space-fixed projection $m_{J, a=1,2}$ along the magnetic-field direction, the spherical harmonics $Y_{\ell m_\ell}(\theta,\phi)$ describe molecular rotation with partial wave $\vec \ell$, and where the angles $\theta$ and $\phi$ orient the internuclear axis relative to the magnetic field.

For a closed-coupling calculation of the rovibrational motion and of the scattering of the atoms we need all electronic potentials dissociating to two ground-state atoms. There are 91 BO potentials for Er$_2$, of which 49 are gerade and 42 are ungerade potentials. For collisions of bosons in the same Zeeman state only gerade states matter. These potential surfaces have been obtained using an {\it ab~initio} relativistic multi-reference configuration-interaction method (RMRCI) \cite{Kotochigova1998crc}, and converted into a tensor operator form with $R$-dependent coefficients. Examples of tensor operators are the exchange interaction $V_{\rm ex}(R) \vec{j}_1 \cdot \vec{j}_2$ and the anisotropic quadrupole-rotation operator $V_{\rm Q}(R) Y_{2}(\hat R)\cdot [ \vec{j}_1 \otimes \vec{j}_1]_2$ coupling the quadrupole operator  $[ \vec{j}_1 \otimes \vec{j}_1]_2$ of one atom with angular momentum $j_1$ to the rotation of the molecule. See \cite{Petrov2012aif} for other operators.

Collisions of submerged 4f-shell atoms at low temperatures also depend on the intermediate to long-range isotropic and anisotropic dispersion, magnetic dipole-dipole and quadrupole-quadrupole interatomic interactions. The van der Waals dispersion potentials for two ground-state atoms are obtained using the transition frequencies and oscillator strengths \cite{nist2013,Lawler2010atp}. The quadrupole moment of Er is calculated using an unrestricted atomic coupled-cluster method with single, double, and perturbative triple excitations uccsd(t) \cite{Watts1993ccm} and shown to be small at $Q=0.029\,\mathrm{a.u.}$. 

We use a first-principle coupled-channel model to calculate anisotropy-induced magnetic Fano-Feshbach-resonance spectra of bosonic Erbium. The model treats the Zeeman, magnetic dipole-dipole, and isotropic and anisotropic dispersion interactions on equal footing. The Hamiltonian includes 
\begin{equation*}
H = -\frac{\hbar^2}{2\mu}\frac{d^2}{dR^2} + 
      \frac{{\vec \ell}^{\,2}}{2\mu R^2} + H_\mathrm{Z} +V(\vec R,\tau)\,,
\end{equation*}
where $\vec R$ describes the orientation of and separation between the two atoms. The first two terms are the radial kinetic and rotational energy operators, respectively. The Zeeman interaction is $H_\mathrm{Z}=g\mu_\mathrm{B} (j_{1z}+j_{2z}) B$, where $g$ is an atomic g-factor and $j_{iz}$ is the $z$ component of the angular momentum operator $\vec\jmath_i$ of atom $i=1,2$. The interaction, $V(\vec R,\tau)$, includes the Born-Oppenheimer and the magnetic dipole-dipole interaction potentials, which are anisotropic, and $\tau$ labels the electronic variables. Finally, $\mu$ is the reduced mass and for $R\to\infty$ the interaction $V(\vec R,\tau)\to0$. Coupling between the basis states is due to $V(\vec R,\tau)$, inducing either isotropic ($\ell$ and $m_\ell$ conserving) or anisotropic ($\ell$ or $m_\ell$ changing) couplings. The Hamiltonian conserves $M_{\rm tot}=m_{J,1}+m_{J,2}+m_\ell$ and is invariant under the parity operation so that only even (odd) $\ell$ are coupled. In the atomic basis set, the Zeeman and rotational interaction are diagonal.

\textbf{NNS probability distribution.}
As the density of resonances is not constant below $30\,\mathrm{G}$ we restrict our analysis to resonances appearing from $30$ to $70\,\mathrm{G}$. We plot a histogram of spacings between adjacent resonances given by $d_{i}=B_{i+1}-B_{i}$. For this an appropriate number of bins is chosen on the order of $\sqrt{N}$, with $N$ being the total number of Fano-Feshbach resonances observed up to $70\,\mathrm{G}$. This ensures a bin size at least an order of magnitude larger than the mean resolution of the trap-loss spectroscopy scan. For every bin a statistical counting error according to a Poisson distribution is assigned. Next, the magnetic-field axis of the histogram is divided by the mean spacing of resonances to get the dimensionless quantity $s=B/\overline{d}$. To calculate the NNS probability distribution $P(s)$ the histogram has to be normalized such that $\int\limits_{0}^{\infty} ds\,P(s) = 1$. As shown in Ref.\,\cite{Brody1981rmp}, the probability distribution of uncorrelated random numbers is simply given by the Poisson distribution $P_{\mathrm{P}}(s) = \mathrm{exp}(-s)$. A theoretical spacing distribution of random matrices can not be written in a simple form but, according to the Wigner surmise, an excellent approximation is given by the Wigner-Dyson distribution $P_{\mathrm{WD}}(s) = \frac{\pi}{2} s\,\mathrm{exp}(-\pi s^2 /4)$. A way of discriminating between these two distributions is to fit the so-called \emph{Brody distribution} to the NNS distribution \cite{Brody1973asm}. It is an empirical function with a single fitting parameter $\eta$, which interpolates between $P_{\mathrm{WD}}$ and $P_{\mathrm{P}}$ and quantifies the tendency (and not the degree of chaoticty) of the observed distribution to be more Poisson-like ($\eta=0$) or more Wigner-Dyson-like ($\eta=1$). It is defined by 
\begin{eqnarray*}
P_{\mathrm{B}}(s) &=& A s^{\eta}\,\mathrm{exp}(-\alpha s^{\eta+1})\\
A &=& (\eta+1)\alpha\\
\alpha &=& \left[ \Gamma \left( \frac{\eta+2}{\eta+1} \right) \right]^{\eta+1}\ ,
\end{eqnarray*}
where $\Gamma$ denotes the Gamma function. From a least-squares fit to the experimental data, we obtain $\eta_{\mathrm{168}}=0.66(10)$ for $^{168}$Er and $\eta_{\mathrm{166}}=0.73(18)$ for $^{166}$Er, and a fit to the CC-calculation data gives $\eta_{\mathrm{CC}}=0.72(18)$.

\textbf{Number variance.}
The number variance $\Sigma^{2}$ is a quantity that depends on long-range correlations between resonance spacings within an interval $\Delta B$. It is defined by
\begin{equation*}
\Sigma^{2}(\Delta B) =\overline{n^{2}(B_{0},\Delta B)}-(\overline{n(B_{0},\Delta B)})^{2}\ ,
\end{equation*}
with $n(B_{0},\Delta B)= \mathcal{N}(B_{0}+\Delta B)-\mathcal{N}(B_{0})$ giving the number of resonances in the interval $[B_{0},B_{0}+\Delta B]$ and the bar denotes the mean value over all $B_{0}$. For a Poisson distribution, $\Sigma^{2}=\Delta B$. By contrast, for a spectrum according to RMT one expects $\Sigma^{2}=1/\pi^{2} \left( \mathrm{ln}(2\pi \Delta B) + \gamma + 1 - \pi^{2}/8 \right)$, for large $\Delta B$ and where $\gamma = 0.5772...$ is Euler's constant \cite{Dyson1963sto}. This behavior reflects that there are only very small fluctuations around an average number of resonances within a given interval of size $\Delta B$ (spectral rigidity). Compared to the NNS distribution the number variance is more suitable to probe long distances in the spectrum. A clear signature of level repulsion on the one hand and a large spectral rigidity on the other are central properties of strong correlations between levels according to RMT \cite{Brody1981rmp}.

\clearpage
%\section*{EXTENDED DATA}

\setcounter{figure}{0} 
\renewcommand{\figurename}{Extended Data Figure}
\makeatletter 
\renewcommand{\thefigure}{\@arabic\c@figure} 
\makeatother

%\textbf{Extended Data}

\begin{figure*}[h]
  \includegraphics[scale=0.95]{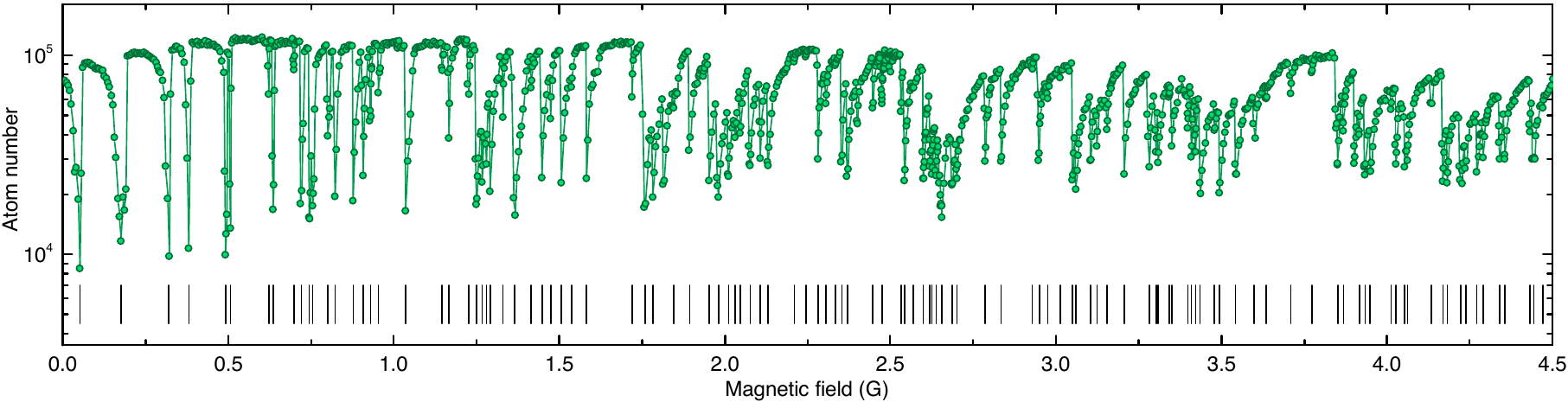}
  \caption{\textbf{Fano-Feshbach spectrum of fermionic $^{167}$Er from $0$ to $4.5\,\mathrm{G}$.} The trap-loss spectroscopy is performed in an optically trapped sample of fermionic Er atoms at a temperature of $0.4\,T_{\rm F}$, where $T_\mathrm{F}=1.0(1)\,\mathrm{\mu K}$ is the Fermi temperature. The atoms are spin-polarized in the lowest Zeeman sublevel, $m_{F}=-19/2$. We keep the atomic sample at the magnetic probing field for a holding time of $100\,\mathrm{ms}$. We observe $115$ resonances up to $4.5\,\mathrm{G}$, which we attribute to be Fano-Feshbach resonances between identical fermions. The corresponding mean density is about $26$ resonances per gauss.}
  \label{figed:spectrfermions}
\end{figure*} 

\begin{figure*}[h]
  \includegraphics[scale=1]{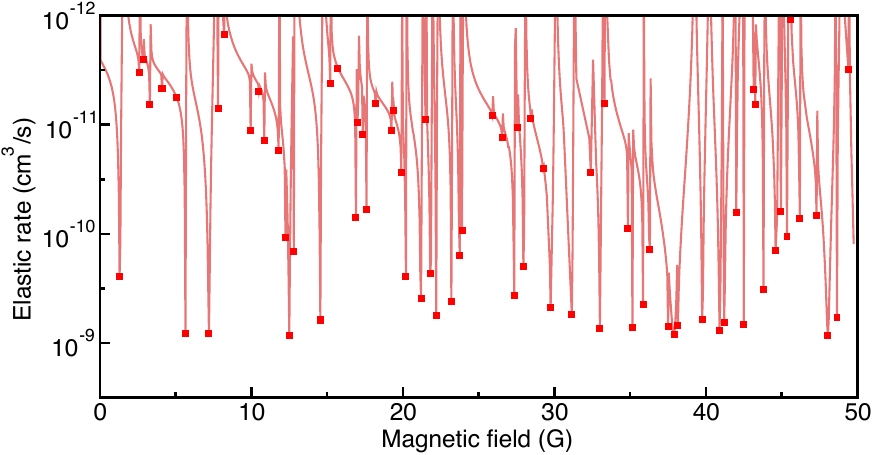}
  \caption{\textbf{Elastic rate coefficient of $m_{J} = -6$ $^{168}$Er collisions.} The $s$-wave elastic rate coefficient as a function of magnetic field assuming a collision energy of $E/k_{\rm B}=$ 360 nK. Partial waves $\ell$ up to 20 are included. A divergence of the elastic rate coefficient, i.e. the position of a Fano-Feshbach resonance, is marked with squares.}
  \label{figed:CCrate}
\end{figure*}   

\begin{figure*}[h]
  \centering
  \includegraphics[scale=1]{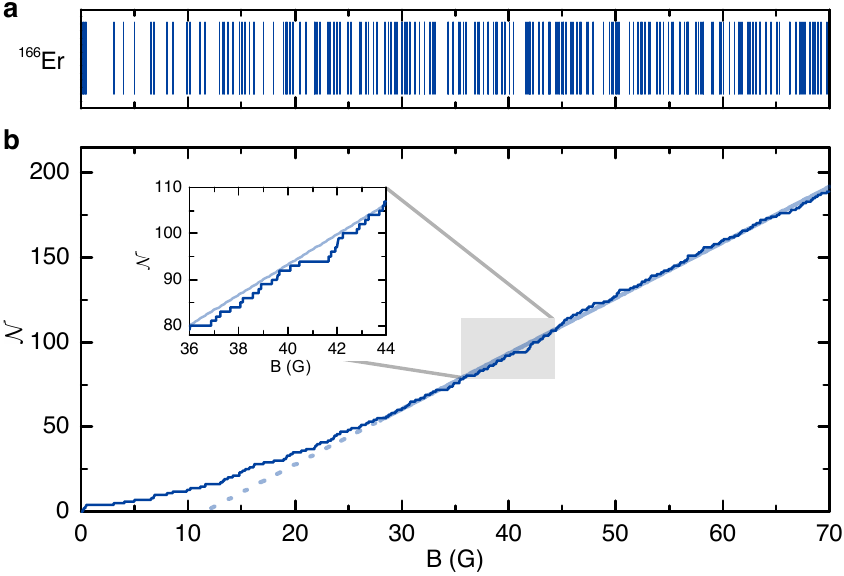}
  \caption{\textbf{Statistical analysis of high-density Fano-Feshbach resonances of isotope $^{166}$Er.} \textbf{a}, Position of the resonances are marked with vertical lines. \textbf{b}, The staircase function shows a similar behavior to $^{168}$Er (Fig.\,\ref{fig:staircase}). A linear fit to the data above $30\,\mathrm{G}$ is plotted in light colors. From the staircase function we calculate a mean density of resonances of $\overline{\rho}=3.3(3)\,\mathrm{G^{-1}}$, which corresponds to a mean distance between resonances of $\overline{d}=0.31(3)\,\mathrm{G}$.}
  \label{figed:statiso}
\end{figure*}

\begin{figure*}[h]
  \centering
  \includegraphics[scale=1]{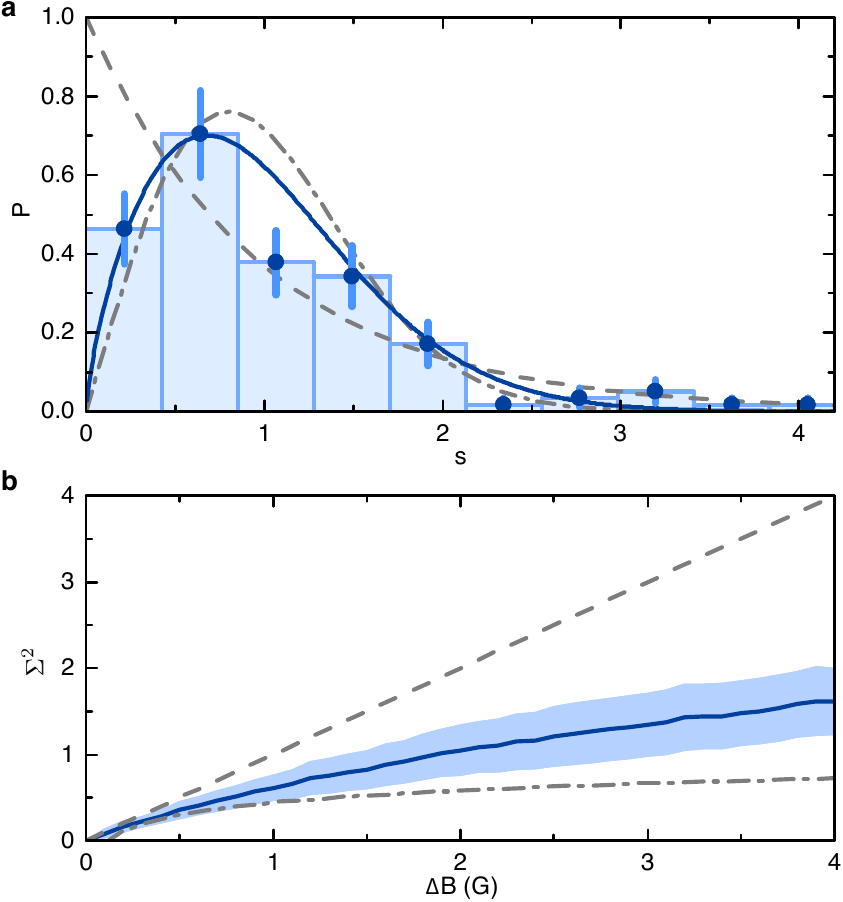}
  \caption{\textbf{NNS distribution and number variance.} \textbf{a}, $^{168}$Er NNS distribution above $30\,\mathrm{G}$ with a bin size of $140\,\mathrm{mG}$. For the error bars we assume a Poisson counting error. The plot shows the experimental data (circles) with the corresponding Brody distribution (solid line). The parameter free distributions $P_\mathrm{P}$ (short-dashed line) and $P_\mathrm{WD}$ are shown and reduced chi-squared values are $\tilde{\chi}_{\mathrm{P}}^2=2.32$ for the Poisson and $\tilde{\chi}_{\mathrm{WD}}^2=1.85$ for the Wigner-Dyson distribution. \textbf{b}, Number variance $\Sigma^{2}$ for the same experimental data (solid line) with a two-sigma confidence band (shaded area). The number variance from experimental data shows a clear deviation from the number variance of a Poisson distribution.}
  \label{figed:distriso}
\end{figure*}

\end{document}